\documentclass[twoside,12pt]{article} 
\usepackage{epsfig} 
 
\def\Journal#1#2#3#4{{#1} {#2} (#4) #3 } 
 
\def\AA{{\em Astron. \& Astrophys.}}

\def\APJ{{\em Astrophys. J.}} 
\def\APJS{{\em Astrophys. J. Suppl.}}

\def\NA{{\em New Astron.}} 
\def\NPA{{\em Nucl. Phys.} A} 
\def\NPB{{\em Nucl. Phys.} B}

\def\PLB{{\em Phys. Lett.} B}

\def\PRL{\em Phys. Rev. Lett.} 
 
\def\PREP{\em Phys. Rep.} 
 
\def\PRD{{\em Phys. Rev.} D} 
\def\PRC{{\em Phys. Rev.} C}

\def\arx{{\em arXiv:}}

\newcommand{\be}{\begin{equation}} 
\newcommand{\ee}{\end{equation}} 
\newcommand{\bea}{\begin{eqnarray}} 
\newcommand{\eea}{\end{eqnarray}} 
 
\newcommand{\lsim}{\stackrel{\scriptstyle <}{\phantom{}_{\sim}}} 
\newcommand{\gsim}{\stackrel{\scriptstyle >}{\phantom{}_{\sim}}} 
\begin{document} 
 
\title{ \vspace{1cm} 
Unmasking neutron star interiors using cooling simulations} 
 
\author{D.\ Blaschke,$^{1,2,3}$ and  H.\ Grigorian,$^{1,4,5}$ 
\\ 
$^1$Institute of Physics, Rostock University, D-18051 Rostock, Germany\\ 
$^2$Institute for Theoretical Physics, University of Wroclaw, 50-204 Wroclaw, 
Poland\\ 
$^3$Bogoliubov Laboratory for Theoretical Physics at JINR, 141980 Dubna, 
Russia \\ 
$^4$Laboratory for Information Technologies at JINR, 141980 Dubna, Russia \\ 
$^5$Department of Physics, Yerevan State University, 375049 Yerevan, Armenia\\ 
} 
\maketitle 
\begin{abstract} 
We introduce a new tool for ``unmasking'' the composition of neutron star (NS) 
interiors which is based on the fact that the state of matter at high densities
determines the statistics of both NS observables, the temperature-age (TA) data
as well as the mass distribution. 
We use modern cooling simulations to extract distributions of NS masses 
required to reproduce those of the yet sparse data in the TA plane. 
By comparing the results with a mass distribution for young, nearby NSs from 
population synthesis we can sharpen two NS cooling problems. 
The {\it direct Urca (DU) problem} consists in a narrowing of the 
NS population at the mass value for which the DU process as the most effective 
cooling mechanism in the hadronic layer of the star can occur. 
The {\it Vela mass problem} is a broadening of the 
population beyond the range of the typical mass window of 1.1 - 1.5 $M_\odot$. 
Applying this tool to modern EoS we discuss examples for pure hadronic stars 
which are in conflict with these constraints while hybrid stars with a 
color superconducting quark matter core can predict a satisfactory mass 
distribution, provided the smallest diquark pairing gap has a properly defined 
density dependence. 
\end{abstract} 
 
\section{Introduction} 
\label{intro} 
 
The quest for understanding the state of nuclear matter at high densities 
beyond saturation at $n\gsim 0.16$ fm$^{-3}$ has triggered a multitude of 
theoretical and experimental investigations. 
It is of principal interest to clarify the mechanism how the quark and gluon 
substructure of nucleons might manifest itself under extreme compression
\cite{Blaschke:2006xt}.
One of the central questions is that of deconfinement: 
nature might choose to excite more massive degrees of freedom of the hadron 
spectrum such as hyperons and nucleon resonances in order to satisfy the Pauli 
principle before the quark-gluon plasma shall emerge as expected from the 
asymptotic freedom property of QCD at ultimate densities. 
Due to the presence of strong attractive correlations in this intermediate 
density regime, the transition of hadrons from well-localized quark bound 
states to delocalized, unbound states is expected to pass a transitory 
phase of strongly correlated few-quark complexes with frequent resonant 
scattering processes \cite{Ropke:1986qs,Barrois:1977xd}.
Thereby a very rich structure of quark condensates can occur characterizing 
color superconducting phases in the low-temperature QCD phase diagram,
most relevant for compact star physics  
\cite{Blaschke:2006xt,Buballa:2003qv,Blaschke:2005uj}.
  
Besides terrestrial experiments with relativistic heavy-ion collisions the 
interior of neutron stars provides a ''laboratory'' where matter under 
conditions of extreme densities occurs \cite{Weber:1999qn}. 
The problem is to identify the composition of neutron star interiors from 
their observable properties like, e.g., masses, radii, rotational and cooling 
evolution \cite{Prakash:1996xs,Blaschke:2001uj}. 
 
Until recently the compactness of objects has been discussed 
as a characteristic feature of stars with a quark matter interior (strange 
stars) when radii are less than $\sim 10$ km \cite{Bombaci:2001uk,Li:1999wt}. 
The situation has dramatically changed with the advent of precision 
measurements of high masses for objects like the pulsar PSR J0751+1807 
\cite{NiSp05} with $M=2.1 \pm 0.2~M_\odot$ or of the mass-radius relation from 
the thermal emission of the isolated neutron star RX J1856-3754 pointing to 
either large radii of $R> 14$ km for a typical NS mass of $\lsim 1.4~M_\odot$ 
or large masses $M\gsim 2.0~M_\odot$ for radii not exceeding 12 km 
\cite{Trumper:2003we}. 
These measurements clearly demand a stiff equation of state and exclude 
standard models for hyperonic or quark matter interiors as well as mesonic 
condensates, see \cite{Ozel:2006bv}. 
 
As has been argued in \cite{Alford:2006vz}, there are several modern 
QCD-motivated quark matter EsoS which could provide enough stiffness of 
high-density matter to be not in conflict with the new mass and mass-radius 
constraints. 
This leads, however, to the effect that hybrid stars with quark matter 
interiors ''masquerade'' as neutron stars \cite{Alford:2004pf} since they 
cannot be distinguished from each other by their mechanical properties, see 
also the contribution by T. Kl\"ahn in this volume 
\cite{Klahn:2006in} for a color superconducting three-flavor 
quark model with selfconsistently determined quark masses and pairing gaps
\cite{Blaschke:2005uj,Klahn:2006iw}. 
 
Therefore, also suggested signals from the timing behavior 
of pulsar spin-down \cite{Glendenning:1997fy}, frequency clustering 
\cite{Glendenning:2000zz} or population clustering 
\cite{Poghosyan:2000mr,Blaschke:2001th} of accreting NSs should not be 
applicable. 
 
In the present work, we suggest a sensitive tool for ''unmasking'' the 
composition of neutron stars which is based on their cooling behavior. 
As the cooling regulators such as neutrino emissivities, heat conductivity and 
specific heat in quark matter might be qualitatively different from those in 
nuclear matter, due to the chiral transition and color superconductivity with 
some possibly sensible density dependence, the TA curves for hybrid stars could
be significantly different from those of neutron stars. 
In order to reach the goal of unmasking the neutron star interior we introduce 
here a new method for the quantitative analysis of the cooling behavior 
consisting in the extraction of a NS mass distribution from the (yet sparse) 
TA data and its comparison with the (most likely) mass distribution 
from population synthesis models of NS evolution in the galaxy 
\cite{Popov:2004ey}. 
 
 
\section{Cooling curves in the TA diagram}

\begin{center} 
\begin{figure}[th] 
\centerline{
\hspace{1cm}\psfig{figure=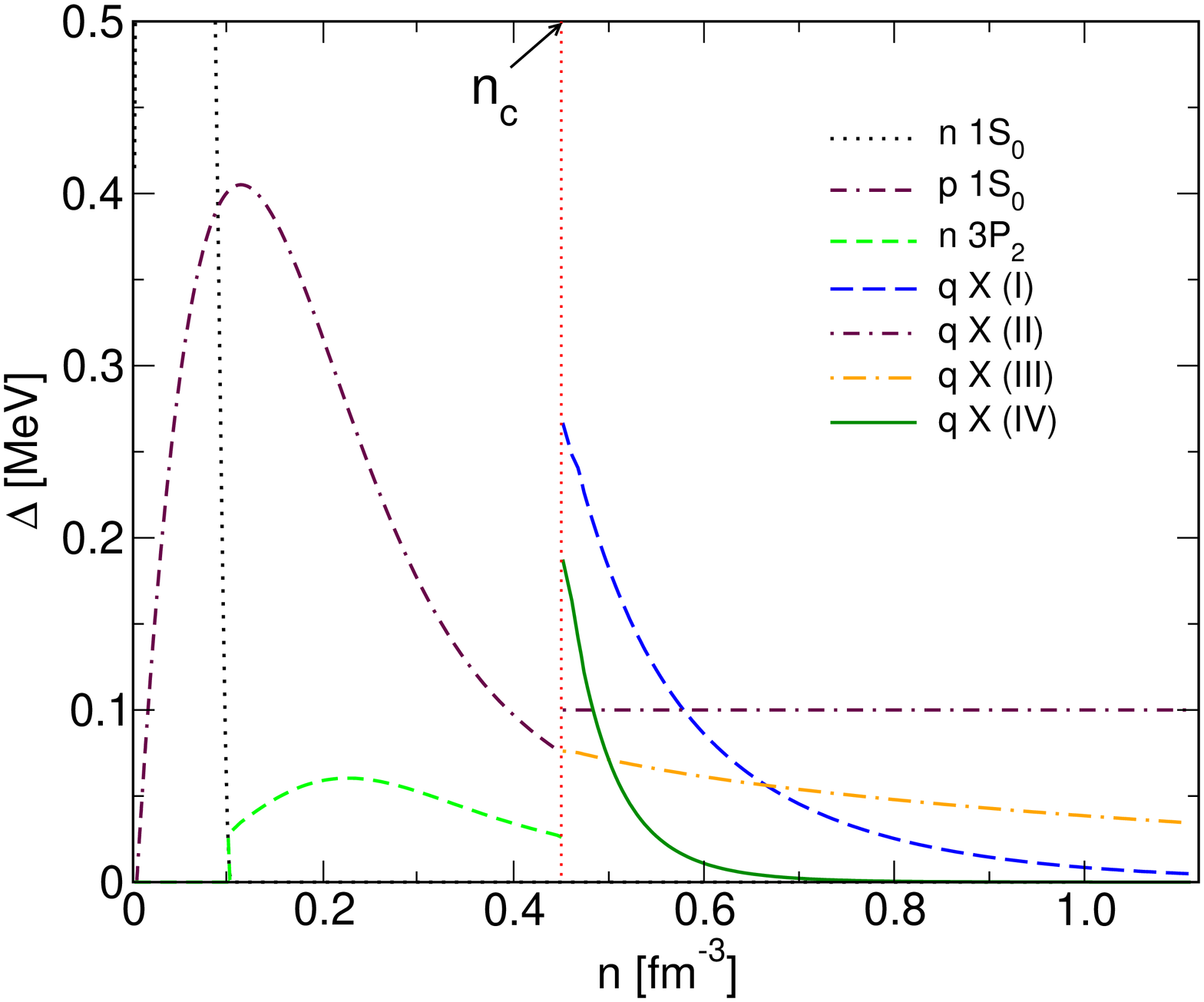,height=10cm,width=8cm,angle=-90} 
\hspace{-5mm}\psfig{figure=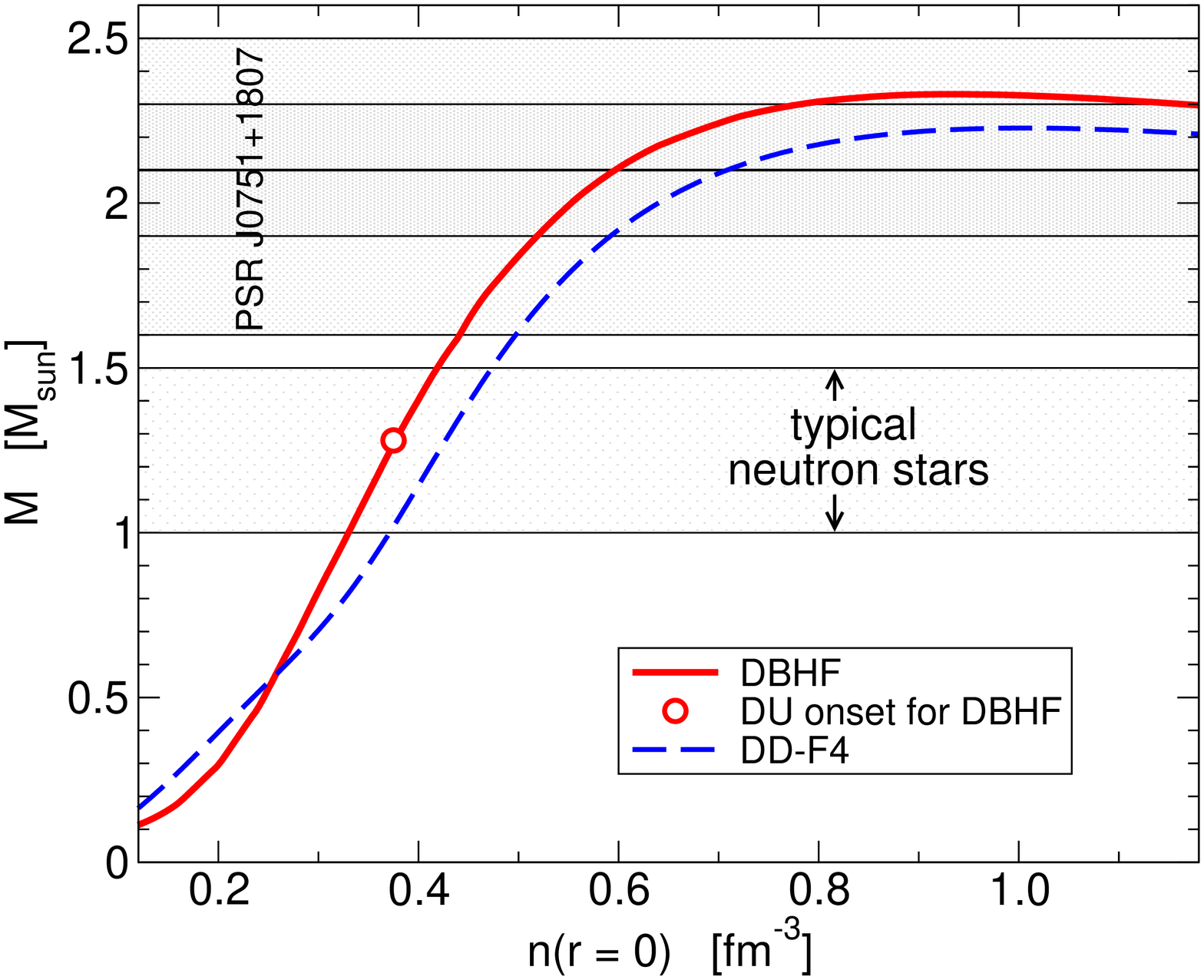,height=10cm,width=8cm,angle=-90}} 
\caption{{\small 
Density dependence of the pairing gaps 
in nuclear matter together with that of the hypothetical X-gap in quark matter 
(left). 
Mass-central density relation for the two 
hadronic EoS models DBHF and DD-F4 (right). 
The dot indicates the onset of the DU process.} 
\label{fig:gaps}} 
\end{figure} 
\end{center} 
 
The cooling behavior of compact stars belongs to the most complex phenomena
in astrophysics. Therefore, the codes for its numerical simulation as
developed by a few groups contain inputs (cooling regulators) of rather 
different kind, see, e.g., 
\cite{Page:2005fq,Page:2004fy,Blaschke:2004vq,Voskresensky:2001fd,Yakovlev:2000jp}. Attempts to develop a {\em Minimal Cooling Paradigm} \cite{Page:2004fy}
by omitting important medium effects on cooling regulators 
\cite{Voskresensky:2001fd,Grigorian:2005fn}
unfortunately result in inconsistencies and suffer therefore from the danger 
of being not reliable. 
To develop a paradigmatic cooling code as an open standard, however, 
is rather necessary to cross-check the present knowledge of the groups
before more sophisticated mechanisms like anisotropies due to the magnetic 
field \cite{Page:2005fq} or special processes in the NS crust or at the 
surface are taken into account. 
Therefore, it is still premature to attempt an identification of the NS 
interior from the cooling behavior. 

In order to circumvent such a model dependence we employ a given cooling code
developed in Refs. \cite{Blaschke:2004vq,Blaschke:2000dy} and vary 
the matter properties such as EoS, superconductivity and star crust model such 
as to fulfill all constraints known up to now (mass, mass-radius, TA, 
brightness, etc.). Moreover, we try to use consistent inputs. 
 
We consider the cooling evolution of young neutron stars with ages 
$t \sim 10^3 - 10^6$ yr which is governed by the emission of neutrinos 
from the interior for $t \lsim 10^5$ yr and thermal photon emission for 
$t \gsim 10^5$ yr. 
The internal temperature is of the order of $T \sim 1$ keV. This is much 
smaller than the neutrino opacity temperature $T_{\rm opac} \sim 1$ MeV as 
well as critical temperatures for superconductivity  in nuclear 
($T_c \sim 1$ MeV) or quark matter ($T_c \sim 1 - 100$ MeV). 
Therefore, the neutrinos are not trapped and the matter is in a
superconducting state. 
In Fig. \ref{fig:gaps} we show the density dependence of the pairing gaps 
in nuclear matter \cite{Takatsuka:2004zq,Blaschke:2004vq} together with that 
of the hypothetical X-gap in quark matter 
\cite{Blaschke:2004vr,Grigorian:2004jq,Popov:2005xa}.
The phase transition occurs at the critical density 
$n_c = 2.75~n_0=0.44$ fm$^{-3}$. 

The main neutrino cooling processes  in hadronic matter are the direct Urca 
(DU), the medium modified Urca (MMU) and the pair breaking and formation 
(PBF) whereas in quark matter the main processes are the quark 
direct Urca (QDU), quark modified Urca (QMU), quark bremsstrahlung 
(QB) and quark pair formation and breaking (QPFB) \cite{Jaikumar:2001hq}.
Also the electron bremsstrahlung (EB), and the massive gluon-photon decay 
(see \cite{Blaschke:1999qx}) are included. 
 
The $1S_0$ neutron and proton gaps in the hadronic shell are taken according 
to the calculations by \cite{Takatsuka:2004zq} 
corresponding to the thick lines in Fig. 5 of  Ref. \cite{Blaschke:2004vq}.  
However, the $3P_2$ gap is suppressed by a factor 10 compared to 
the BCS model calculation of \cite{Takatsuka:2004zq}, consistent with 
arguments from a renormalization group treatment of nuclear pairing 
\cite{Schwenk:2003bc}. 
Without such a suppression of the $3P_2$ gap the hadronic cooling 
scenario would not fulfill the TA constraint, see \cite{Grigorian:2005fn}. 
 
The possibilities of pion condensation and of other so called 
exotic processes are included in the calculations for purely hadronic stars 
but do not occur in the hybrid ones since the critical density for pion 
condensation exceeds that for deconfinement in our case \cite{Blaschke:2004vq}.
While the hadronic DU process occurs in the DBHF model EoS for all neutron 
stars with masses above $1.27~M_\odot$, it is not present at all in the 
DD-F4 model, see the right panel of Fig.~\ref{fig:gaps}.  
We account for the specific heat and the heat conductivity of all existing 
particle species contributing with fractions determined by the $\beta-$ 
equilibrium conditions. 
Additionally, in quark matter the massless and massive gluon-photon modes also 
contribute. 
 
In the 2SC phase only the contributions of quarks forming Cooper pairs 
(say red and green) are suppressed via huge diquark gaps,
while those of the remaining unpaired blue color lead to a so fast cooling 
that the hybrid cooling scenario becomes unfavorable \cite{Grigorian:2004jq}. 
Therefore, we assume the existence of a weak 
pairing channel such that in the dispersion relation of 
hitherto unpaired blue quarks a small residual gap can appear. 
We call this gap $\Delta_X$ and show that for a successful description 
of the cooling scenario $\Delta_X$ has to have a density dependence. 
We have studied the ansatz  
\begin{equation} 
\Delta_{\mathrm{X}}= \Delta_0 \, \exp{\left[-\alpha\, \left( 
\frac{\mu - \mu_c}{\mu_c}\right)\right]} \label{gap}~, 
\end{equation} 
where $\mu$ is the quark chemical potential, $\mu_c=330$ MeV. 
For the analyses of possible models we vary the values of 
$\alpha$ and $\Delta_0$, given in the Table 1 of \cite{Popov:2005xa} and 
shown in the left panel of Fig. \ref{fig:gaps}. 
 
The physical origin of the X-gap is not yet identified. It could 
occur, e.g., due to quantum fluctuations of color neutral quark 
{sextet} complexes \cite{Barrois:1977xd}. Such calculations have 
not yet been performed within the relativistic chiral quark models. 
The size of the small pairing gaps in possible residual single color/ 
single flavor channels \cite{Schafer:2000tw} is typically in the interval 
$10~$ keV - $1~$MeV, see discussion in \cite{Alford:2002rz}. 
The specific example of the CSL phase is analyzed in  more in detail in
Refs. \cite{Aguilera:2005tg,Schmitt:2005wg,Aguilera:2006cj}. 
 
\begin{center} 
\begin{figure}[th] 
\centerline{
\hspace{1cm}\psfig{figure=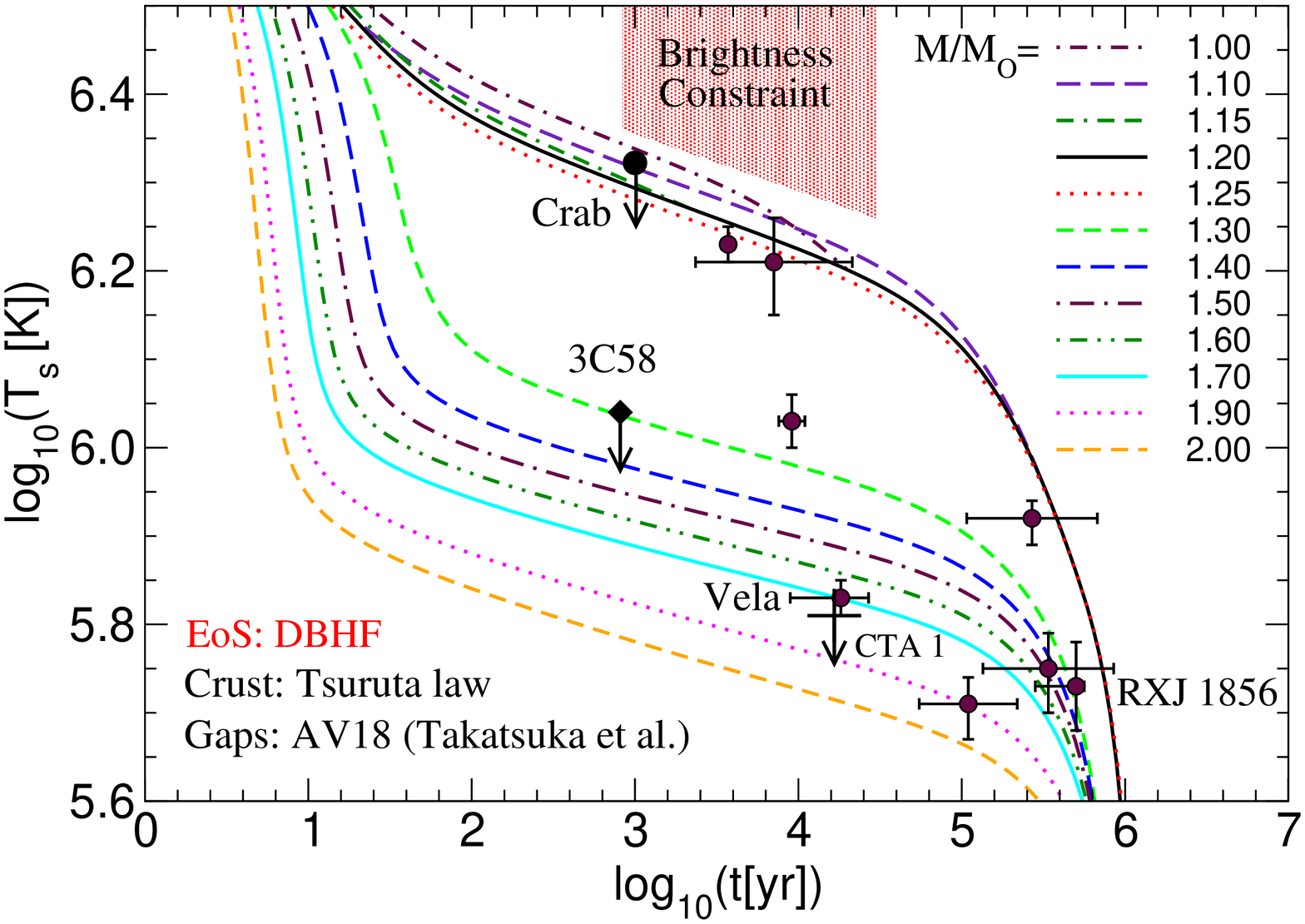,height=9.5cm,width=8cm,angle=-90} 
\hspace{-5mm}\psfig{figure=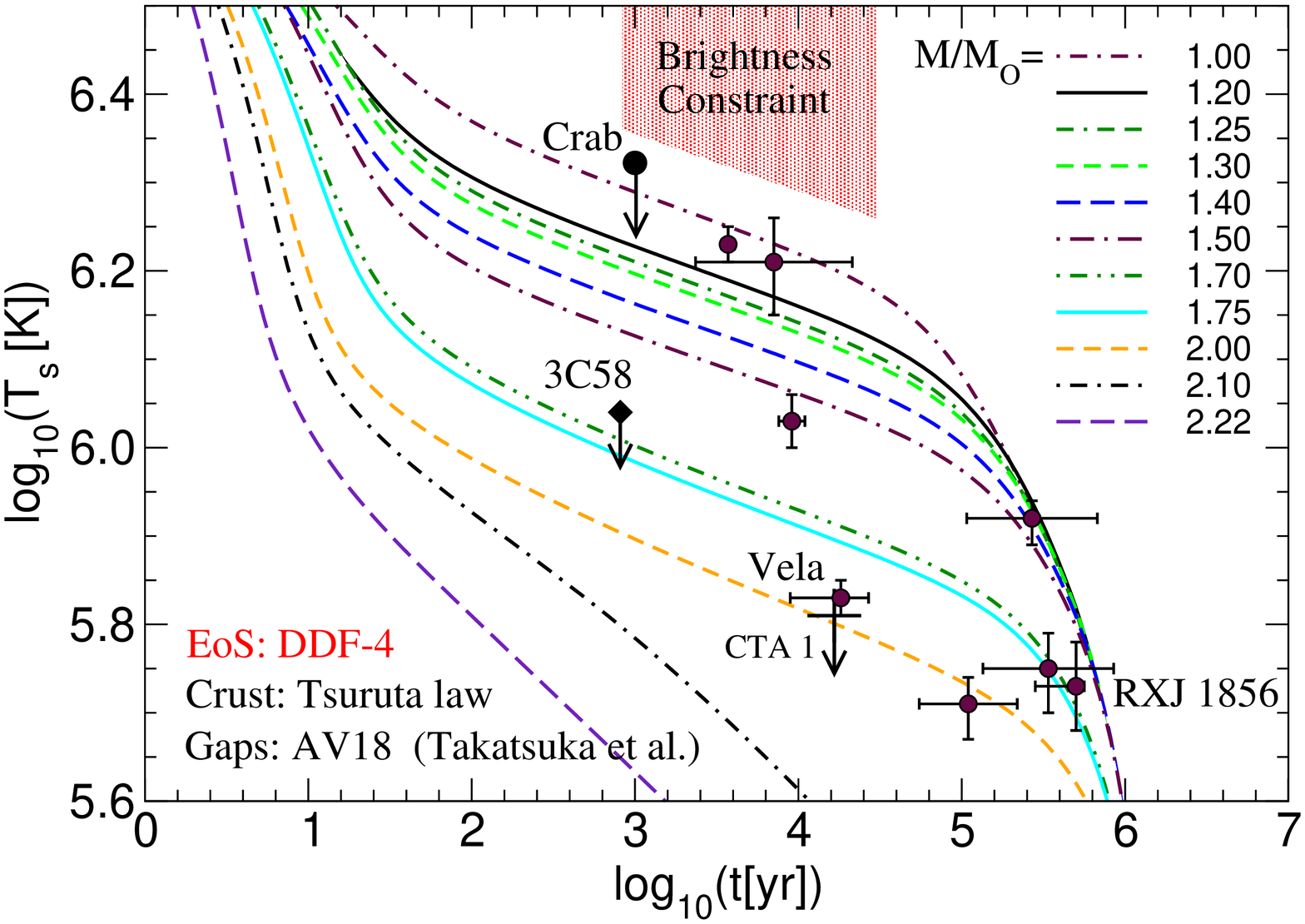,height=9.5cm,width=8cm,angle=-90}} 
\caption{{\small Hadronic star cooling curves for DBHF model EoS. 
Different lines correspond to compact star mass values indicated 
in the legend (in units of $M_\odot$), data points with error bars 
are taken from Ref. \cite{Page:2004fy}.} 
\label{fig:hc1}} 
\end{figure} 
\end{center} 
 
In Fig.~2 we present TA-diagrams for two hadronic star EsoS, where 
DBHF is an ab-initio calculation for the Bonn-A nucleon-nucleon potential 
within the Dirac-Brueckner-Hartree-Fock approach \cite{DaFuFae05}, discussed 
in the context of compact star constraints in 
\cite{Klahn:2006ir,Grigorian:2006pu}. 
DD-F4 denotes a relativistic mean-field model of the EoS 
with density-dependent masses and coupling constants adjusted to mimick the 
behavior of the DBHF approach \cite{Typel:2005ba,Typel}. 
In the hadronic cooling calculations presented in Fig.~2 the crust model has 
been chosen according to Tsuruta's formula for the $T_m-T_s$ relationship 
between be temperatures  of the inner crust and the surface 
\cite{Blaschke:2004vq}.

In Fig.~3 we show the TA diagrams for two hybrid star cooling models presented 
in Ref. \cite{Popov:2005xa}. 
The TA data points are taken from \cite{Page:2004fy}. 
The hatched trapeze-like region represents the brightness 
constraint (BC) \cite{Grigorian:2005fd}. 
For each model nine cooling curves are shown for configurations with mass 
values corresponding to the binning of the population synthesis 
calculations explained in \cite{Popov:2005xa}. 
 
\begin{center} 
\begin{figure}[th] 
\centerline{
\hspace{1.5cm}\psfig{figure=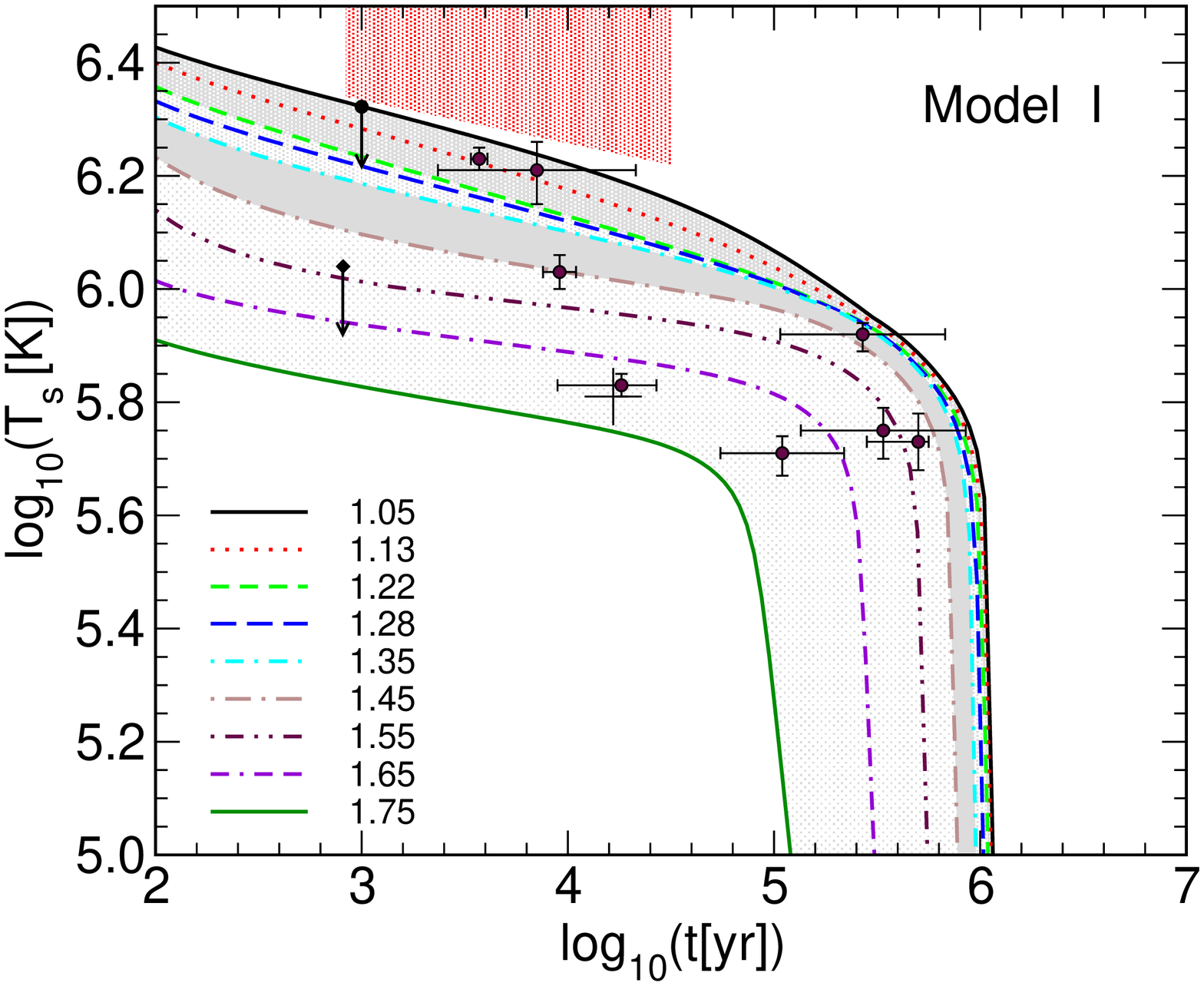,height=11cm,width=8cm,angle=-90} 
\hspace{-1.5cm}\psfig{figure=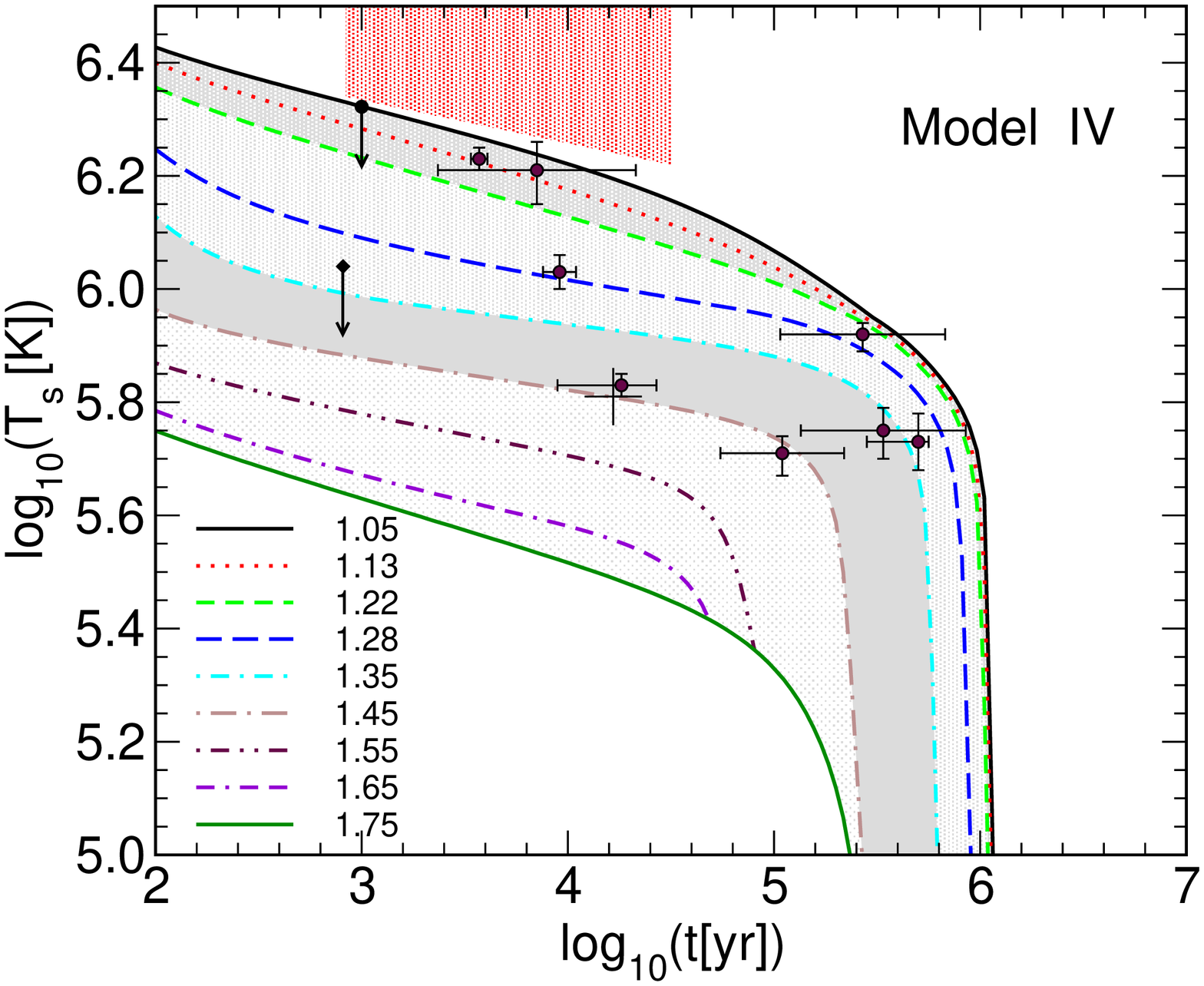,height=11cm,width=8cm,angle=-90}} 
\caption{{\small Cooling curves for hybrid star configurations with 2SC+X 
pairing pattern and X-gap model I (left) versus model IV (right). For the gaps
see the left panel of Fig.~1. The grey value for the shading of the mass bin 
areas corresponds to the probability for that mass bin value in the population 
synthesis model of Ref. \cite{Popov:2004ey}.} 
\label{fig:qc1}} 
\end{figure} 
\end{center}

 
For the hybrid cooling scenario in \cite{Popov:2005xa} a more 
detailed measure for the ability of a cooling model to describe 
observational data in the temperature-age diagram had been 
introduced. 
Also the logN-LogS distribution constraint has been considered. 
In the TA diagram to encode the likelihood that 
stars in that mass interval can be found in the solar 
neighborhood, in accordence with the population synthesis 
scenario, see Fig. \ref{fig:qc1} we use a marking with five grey 
values. The darkest grey value, for example, corresponds to the 
most populated mass interval $1.35$ - $ 1.45~M_\odot$ predicted 
by the mass spectrum used in population synthesis.

 

 
\section{Mass distribution from TA data}

 
In the present work we suggest a method to construct a NS mass distribution 
as a tool for the quantitative characterisation of a given cooling model 
using the yet sparsely distributed measurements of cooling neutron stars in 
the TA plane as a probability measure. 
This method is described as follows. 
Once the cooling model is defined by the EoS and the cooling regulators, 
the configuration corresponding to a gravitational mass $M$  
and its cooling curve $T(t;M)$ can be determined. 
Next a set of mass values $M_i$, $i=0,\dots,N_M$ is introduced, defining the 
borders of $N_M$ mass bins as, for example, in the population synthesis.
A pair of neighboring cooling curves $T(t;M_i)$ and  $T(t;M_{i-1})$ 
corresponds to the $i^{th}$ mass bin and delimits a strip in the TA plane.
We now construct a measure for the number of cooling objects to be expected
within this mass bin 
\begin{equation}
N_i=\sum_{j=1}^{N_{\rm cool}}\int dt\int_{T(t;M_i)}^{T(t;M_{i-1})} dT P_j(T,t),
\label{Ni}
\end{equation}
where $N_{\rm cool}$ denotes the total number of observed coolers used for the 
analysis and $P_j(T,t)$ is the probability density to find the $j^{th}$ object 
at the point $(T,t)$ in the TA plane. 
For the present exploratory study we make the simplest ansatz that  $P_j(T,t)$
is constant in the rectangular region defined by the upper and lower limits
of the confidence intervals corresponding to the temperature and age 
measurements, $(T_{jl},T_{ju})$ and $(t_{jl},t_{ju})$, respectively,
\begin{equation}
P_j(T,t)=[(t_{jl}-t_{ju})(T_{jl}-T_{ju})]^{-1}\Theta(T-T_{jl})\Theta(T_{ju}-T)
\Theta(t-t_{jl})\Theta(t_{ju}-t)~.
\label{Pj}
\end{equation}
Note that in the case when the exact age the object is known
(e.g., for a historical supernova), the time-dependence of  $P_j(T,t)$ 
degenerates to a $\delta$-function and the $t$-integral in (\ref{Ni}) can be 
immediately carried out, leaving us with a one-dimensional probability measure.

This method has been applied to the cooling models for hadronic and hybrid 
stars described in the previous section. The results for the extracted mass 
distributions are normalized to 100 objects, defining 
$N(M)=100 ~ N_i/(\sum_{i=0}^{N_M} N_i)$, and shown in Fig.~4.   
 
 
As we see from Fig.~4, the results are very sensitive to the chosen cooling 
model. 
In the hadronic scenario the onset of the DU cooling mechanism drastically
narrows the mass distribution around the critical mass for the DU onset, see 
Fig.~1. 
On the other hand the slow cooling model predicts more massive objects 
than could be justified from the independent population analysis. 
 
When comparing the density dependence of the pairing gaps, given in the left 
panel of Fig.~1, with the extracted mass distributions for the corresponding 
hybrid models in the right panel of Fig.~4, the direct relationship between 
the superconductivity and the mass distribution becomes obvious. 
 
The DU problem as it was previously discussed in the literature 
\cite{Blaschke:2004vq,Klahn:2006ir,Kolomeitsev:2004ff} was based on the 
intuitive understanding that the mass distribution can not be peaked at a 
critical mass value which accidentally is unique for all observed young 
objects. 
Our modification of the definition of the DU problem does not contradict that 
suggestion, but rather provides an additional measure which rehabilitates the 
validity of cooling scenarios including the DU process. 
 
On the other hand, the EoS model should obey the mass constraints too. 
Therefore, using the models discussed in this work we demonstrate that the 
most preferable structure of the compact object is likely to be a hybrid star 
with  properly defined color superconductivity of the quark matter state in 
the core. 
 
\begin{center} 
\begin{figure}[h!] 
\centerline{ 
\hspace{1cm}\psfig{figure=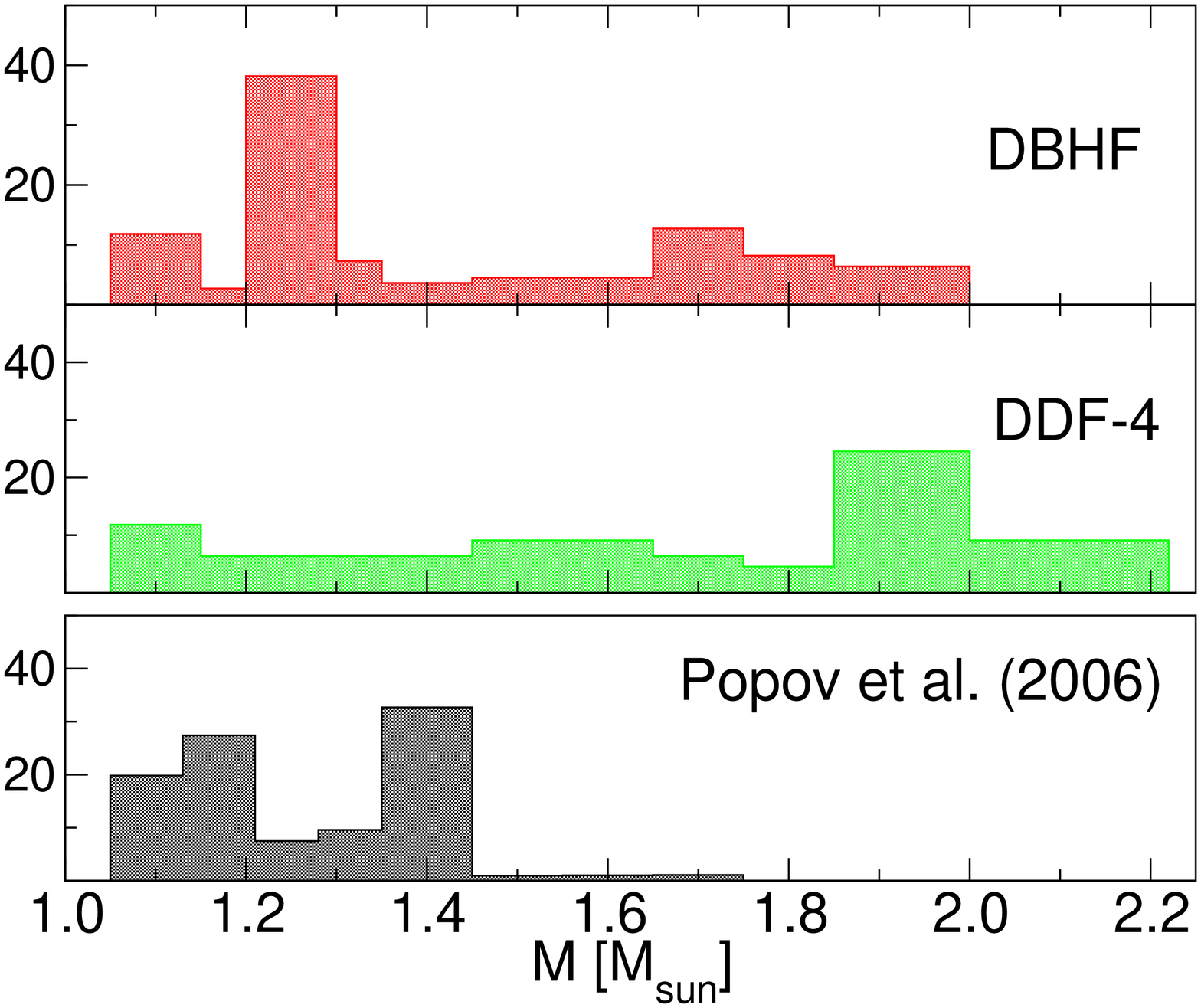,height=10cm,width=8cm,angle=-90} 
\hspace{-5mm}\psfig{figure=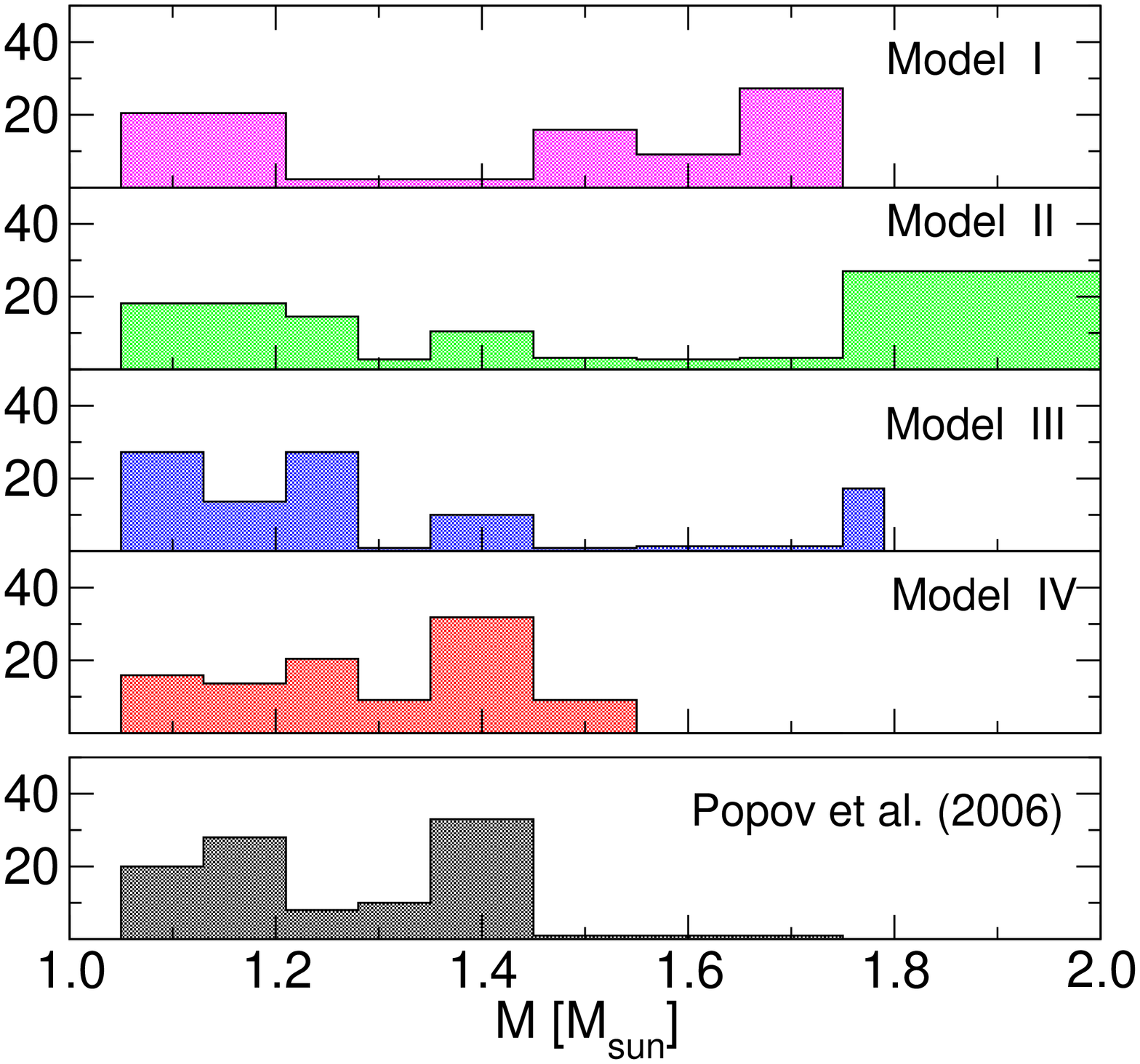,height=11cm,width=8cm,angle=-90}} 
\caption{{\small 
NS mass spectra extracted from the distribution of cooling data 
for both hadronic EoS models (left panel) and for  hybrid stars with 
X-gap models I-IV (right panel).
For comparison, the mass disctribution of young, nearby NS from the 
population synthesis of Popov et al. \cite{Popov:2005xa} is shown at the 
bottom of the panels.
} 
\label{fig:masshb}} 
\end{figure} 
\end{center} 
 
 
\section{Conclusion} 
 
In this contribution we have presented mass distributions obtained by analysing
cooling calculations for NSs and demonstrated that they provide a sensible 
measure for the composition of compact star matter, even if the 
mechanical properties of the compact objects are almost identical. 
The comparison with a mass distribution from a population synthesis approach 
allows to favor hybrid stars with properly defined color superconducting quark 
matter core over other hybrid star or pure neutron star models. 
The presented approach makes testable predictions and will be quantitatively 
improved when a better statistics for TA data will emerge from present and 
future observational programmes. 
On the other hand, an improvement of the population synthesis might be required
in connection with the ongoing development of supernova modeling
and a deeper understanding of astrophysical processes in the galactic 
neighborhood \cite{Popov:2006ki}. 
The new method presented here to extract a NS mass distribution from the 
cooling behavior has proven useful as a new tool to unmask the NS interior 
since it is sensitive to subtle changes in the cooling regulators. 
It shall therefore be further developed and not be missed out when Astronomy 
meets QCD \cite{Popov:2006ia} and more stringent tests are applied to the 
behavior of nuclear matter under conditions of extreme densities in neutron 
stars, a place where the nature of the deconfinement transition can be studied.
 
\section*{Acknowledgements} 
We thank S. Typel for providing the DD-F4 EoS prior to publication.
J. Berdermann and T. Kl\"ahn are acknowledged for their critical reading of 
the manuscript and useful comments.  
D.B. is grateful for the stimulating discussions at the Erice school, 
in particular with K. Langanke, F. Thielemann, J. Wambach, F. Weber and H. 
Wolter. 
This work has been partially supported by the Virtual Institute VH-VI-041 
of the Helmholtz Association for ``Dense hadronic Matter and QCD Phase 
Transition''. 
 


\end{document}